# Experimental and numerical study of shock wave propagation in water generated by Pulsed Arc Electrohydraulic Discharges

*W. Chen[1], O. Maurel[1], C. La Borderie[1], T. Reess[2], A. De Ferron[2], M. Matallah[4], G. Pijaudier-Cabot[3], A. Jacques[5], F. Rey-Bethbeder[5]*

[1] *Sciences pour l'ingénieur en mécanique et génie électrique (SIAME), Université de Pau, Allée du Parc Montaury, 64600 Anglet, France*
[2] *Sciences pour l'ingénieur en mécanique et génie électrique (SIAME), Université de Pau, Hélioparc Pau-Pyrénées, 2 Avenue du Président Angot, 64053 Pau, France*
[3] *Laboratoire des Fluides Complexes et leurs Réservoirs, Université de Pau, BP 1155, 64013 Pau Cedex, France*
[4] *RISAM (RISk Assesment & Management), Université de Tlemcen, BP230, Algérie*
[5] *Total, CSTJF, Avenue Larribau, 64018 Pau Cedex, France*

**Abstract**

The objective of this study is to simulate the propagation of the shock wave in water due to an explosion. The study is part of a global research program on the development of an alternative stimulation technique to conventional hydraulic fracturing in tight gas reservoirs aimed at inducing a distributed state of microcracking of rocks instead of localized fracture. We consider the possibility of increasing the permeability of rocks with dynamic blasts. The blast is a shock wave generated in water by Pulsed Arc Electrohydraulic Discharges (PAED). The amplitude of these shock waves is prescribed by the electrohydraulic discharges which generate high pressures of several kilobars within microseconds. A simplified method has been used to simulate the injected electrical energy as augmentation of enthalpy in water locally. The finite element code EUROPLEXUS is used to perform fluid fast dynamic computation. The predicted pressure obtained in the simulation is consistent with the experimental results.

## 1. Introduction

Tight gas reservoirs contain gas stored in highly impermeable rocks (permeability in the range of 0.1 mD). Conventional oil and gas stimulation techniques use hydraulic fracturing in the periphery of the well to increase the permeability of rocks. Unfortunately, classical hydraulic fracturing produces few localized major cracks and it is only from a small region around those cracks that gas may be extracted. It follows that the production of gas decreases severely with time and that the efficiency of classical hydraulic fracturing remains quite moderate.

The study presented in this paper is part of a global research program on the development of an alternative stimulation technique aimed at inducing a distributed state of microcracking of rocks instead of localized fracture. In quasi-brittle materials such as rocks, micro cracks generated by dynamic load are more distributed compared to those generated with static loads [CAO 01]. Our objective is to benefit from this observation and to consider the possibility of increasing the permeability of rocks with dynamics loads. The dynamic load is a compression shock wave generated by Pulsed Arc Electrohydraulic Discharges (PAED) in water [CHI 98] [KIR 99] [TOU 03]. The amplitude of the pressure wave is controlled by the electrical energy, and by the distance between the electrodes and the surface of specimen in the experiments [TOU 06]. The shock wave is propagated in water and transmitted to the specimen, which is immersed in water [ZHA 05]. An experimental program has been carried out on mortar specimens, in order to observe the effect of a compression pressure wave on permeability [MAU10].
In order to characterize the effects of the shock wave on the host rock, it is necessary to know the time-varying pressure profiles at selected locations both in the fluid and in the rock mass.

This paper discusses the simulation of the generation and propagation of shocks waves in water due to PAED. In the experiments and in the envisioned process on real scale, the underwater shock wave is generated by a dielectric breakdown in a water gap. This breakdown mechanism involves a dynamic process in which an electron avalanche develops in vapour following the formation of a bubble in the water [4]. These vapour bubbles are formed by heating due to the injection of energy into the water. The volume of the bubbles is directly related to the energy injected into the water. When the bubble volume can fill the whole inter-electrode space, dielectric breakdown occurs. The generation of a pressure wave is associated with this breakdown [5].

Over the past decades, hydrodynamic waves produced by electrical breakdown of water have been studied theoretically and experimentally on several occasions. The measurement and analysis of shockwaves generated by pulsed discharge in water made with FUJI pressure measuring film is reported in Ref. [1]. The amplitude of these pressure waves generated underwater by electrical explosion of wires is measured by a piezoelectric pressure probe and can be analysed by a signal processing algorithm which is based on energy conservation requirements and Fourier transform [2]. Meanwhile, the influence of water conductivity on the efficiency of electro-hydraulic shock waves has been considered [3]. A specific study was conducted by TOUYA et al. [4] to characterize the pressure wave based on a number of experimental parameters. An inter-electrode apparatus was used in this study. The pressure waves have been measured with a high frequency piezoelectric sensor, whose response time ranges in the order of nanoseconds. An empirical law was obtained relating the amplitude of the pressure wave to the available electrical energy at breakdown time and to the distance between the arc and the point at which the pressure wave is characterised.

The simulation of shock-wave generation in water by electrical discharges has also been developed in recent years. S. Madhavan et al. [5] performed one-dimensional and two-dimensional hydrodynamic simulations for studying shock production in water. Simulation of blast phenomena in underground electrical power plants involve shock wave generation [6] and were carried out with the Plexis-3C code, jointly developed by the French Commissariat at Energie Atomique (CEA) and the Joint Research Centre of the European Commission (JRC).

In this paper, a simplified model is proposed in order to simulate underwater explosion by electrohydraulic discharges. Our main objective is to develop a calculation method able to provide a correct characterization of shock wave propagation in water (and further in the rock masses around). For simplicity, the details involved in the development and expansion of the plasma channel are not described. The hydrodynamic waves result from the fast injection of energy into water. This (electrical) energy is injected as an augmentation of enthalpy. Thermodynamic equilibrium is assumed, water is described as a two-phase mixture (vapour and liquid) and the simulation uses tabulated equation of state for water. Numerical results are compared with experimental data and consistent agreements are obtained.

## 2. Description of the experimental program

Experiments based on the set-up devised initially by Reess et al. [4] have been conducted in order to provide data against which the simplified model discussed next will be compared. The configuration is presented in figure 1. A point-point electrode system was immersed in a glass fibre tank filled with tap water at room temperature. The electrodes were made of two vertical cylindrical tubes on the lower ends of which were screwed two stainless steel ovoid poles. The shock waves have been recorded with a high frequency piezoelectric sensor which was co-developed by D. Bauer and the Franco-German Research Institute of Saint-Louis (France). During the test, the time histories of the pressure were recorded for several sensor locations.

| Test | Energy (J) | Distance between the sensor and the electrodes (cm) | Inter-electrode distance (mm) |
|---|---|---|---|
| 1 | 3300 | 17.5 | 10 |
| 2 | 600 | 17.5 | 5 |
| 3 | 31 | 9 | 5 |
| 4 | 20 | 9 | 5 |

Table 1. Test parameters

Four tests have been carried out. The test parameters were the injected energy, the distance between the sensor and the electrodes and the inter-electrode distance. The characteristics of the test parameters are illustrated in the table 1.

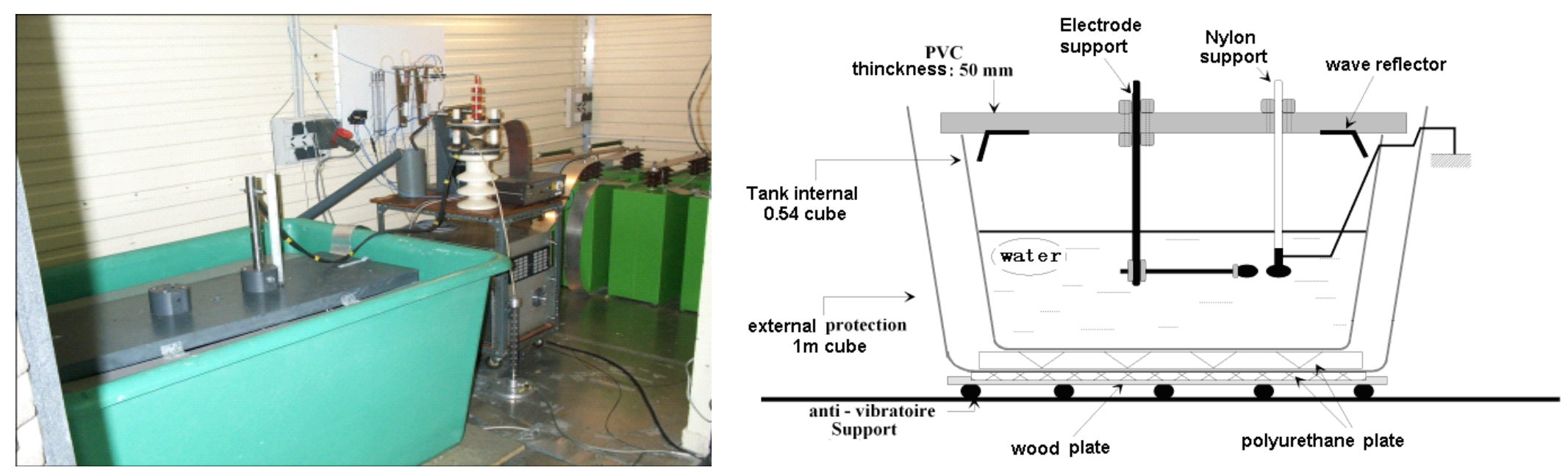

*Figure 1111. Experimental arrangement and test configuration*

In these experiments, electrohydraulic discharges are concerned with thermal processes and water breakdown phenomena, which are subjected to relatively low but long electric field stresses (several tens microsecond duration). Vapor bubbles are formed by volume heating due to the thermal energy. Then, an electron avalanche develops in the vapor phase following the formation the bubbles.

The injected energy at time *t* is calculated by the following relationship:

$$E(t) = \frac{1}{2}.C.(U_M^2 - U(t)^2) \qquad (1)$$

where $U_M$ is the maximum applied voltage, $U_{(t)}$ the voltage level and C the charging capacitor value. When the injection time is long, the bubble volume fills the whole inter-electrode space and gap breakdown occurs, shock waves are generated and propagate in the water. The thermal energy ΔE reads:

$$\Delta E = m.c.\Delta T \qquad (2)$$

*m* is the mass of water, c is the specific heat and ΔT is temperature rise. The previous study [4] provides An empirical formula which relates the peak pressure $P_0$ by the injected energy, the inter-electrode geometry and the distance between the pressure sensor and the plasma channel was calibrated in ref. [4] :

$$P_0 = k.E_B^{\alpha} \qquad (3)$$

where $E_B = 1/2.C.U_B^2$ ($U_B$ is the breakdown voltage value), k and α are parameters which depend on the inter-electrode geometry and the distance d between the pressure sensor and the plasma channel. Comparisons between the numerical model and experimental data, the peak pressure especially are going to be discussed in section 4. Before that, let us describe the numerical model.

## 3. Numerical model

A fast dynamics computer code (EUROPLEXUS) has been used. This finite element code has been implemented on several transient problems such as impact on concrete vessels [13], hydrodynamics [14] [15] [16], pipe whipping [17], and Core Disruptive Accident modelling in nuclear vessels [9-12]. Fluids and solid responses are computed. In the following, the description of the fluid phase is going to be detailed only.

### *3.1 Description of the fluid phase*

The simulation aims at describing electrohydraulic discharge and the associated shock wave propagation in water. Liquid water and vapor are described as a homogeneous mixture, which is associated to tabulated thermodynamic properties of liquid water and vapor (specific volume, enthalpies, heat capacity, saturation). The two phases are assumed to be in thermodynamic equilibrium, with the same temperature and pressure. According to the finite element discretization, properties are characterized by their values computed at the integration points of each finite element. In the present case, each finite element has a single (Gauss) integration point. The underwater explosion due to an electrical discharge is described by the energy released in liquid water as a function of time. At each time step, e.g. between time $t^{(n+1)}$ and time $t^{(n)}$, the increment of the prescribed injected energy Δe is equal to the increment of enthalpy *Δh* of the fluid. At state n +1, the enthalpy is estimated:

$$h^{n+1} = h^n + \Delta h \qquad (4)$$

The thermodynamic properties of the fluid are obtained from tabulated functions of the temperature T and pressure P [Citer : Lester Haar, John Gallagher and George Kell,NBS/NRC Steam TablesThermodynamic and Transport Properties and Computer Programs for Vapor and Liquid States of Water in SI Units,Hemisphere Publishing Corporation, Washington, 1984] :

$$v = v\,(T, P) \quad h = h\,(T, P) \quad s = s\,(T, P) \quad C_p = C_p\,(T, P) \qquad (5)$$

where *v* is the specific volume, *h* is the specific enthalpy, *s* the specific entropy, $C_p$ the specific heat.

For any arbitrary set of pressure and temperature (*T,P)* the quantities are linearly interpolated between the tabulated values.

The liquid and vapor phases of water are separated by the saturation curve depicted in Fig. 2. In the computational model this curve has been also tabulated.

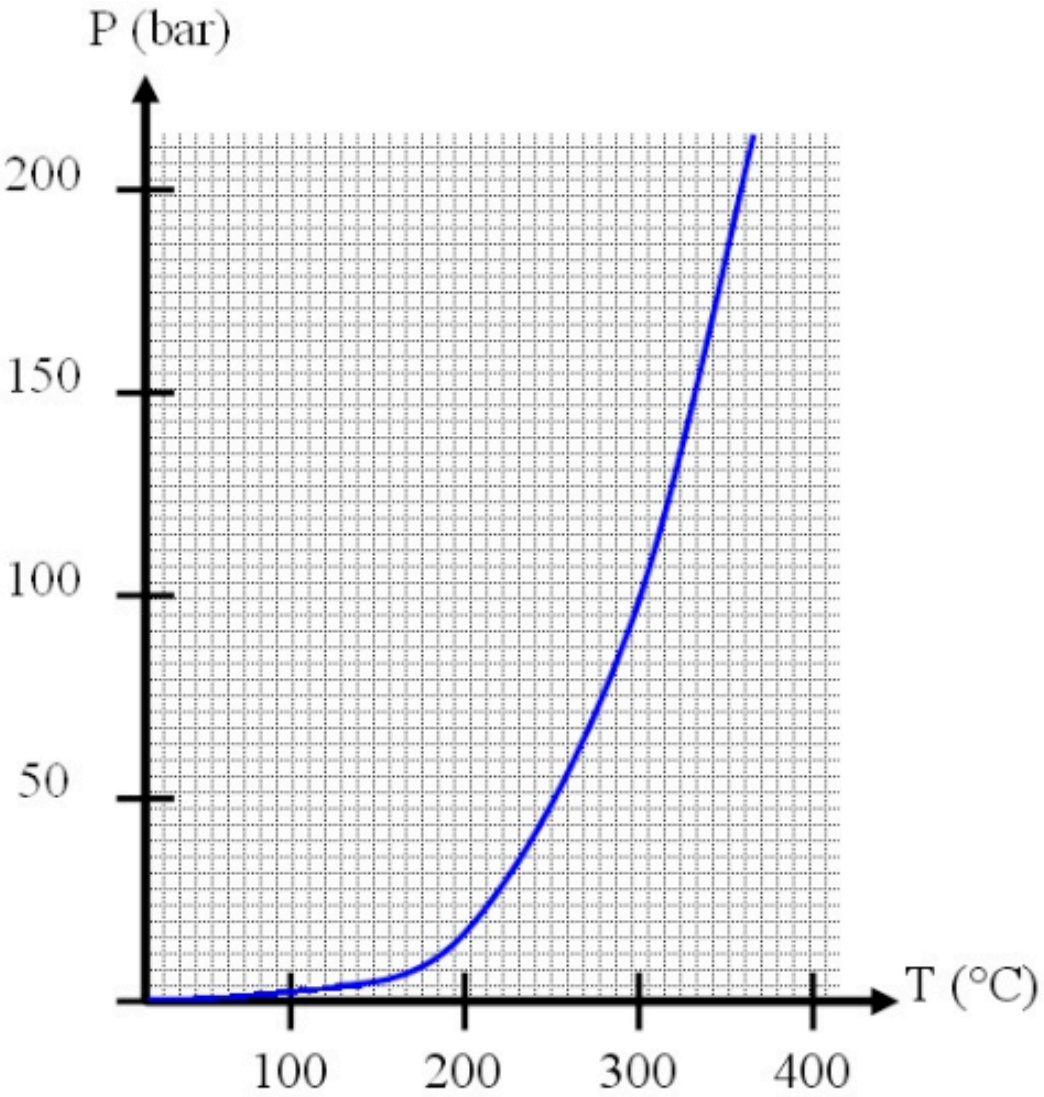

Figure 2: Saturation curve for water. Above the curve vapour is found, below is the domain of liquid water and on the line, vapour and liquid are mixed

In the case of a mixture of liquid water and vapour, the temperature and pressure satisfy the formula which describes the saturation curve :

$$Sat\ (T,\ P)=0 \tag{6}$$

and a new variable is introduced: the mass fraction of vapour $\chi$ :

$$\chi = \frac{m_g}{m} \tag{7}$$

where $m_g$ is the mass of vapour and $m$ is the total mass of fluid. This variable expresses the concentration of vapor. When $\chi$ = 1, the fluid is vapor, when = 0, the fluid is liquid. When 0 < $\chi$ <1, the fluid is diphasic, the mass volume $v$ and mass enthalpy $h$ are given by:

$$v = \chi v_v + (1-\chi)v_l \tag{8}$$

$$h = \chi h_v + (1-\chi)h_l \tag{9}$$

$h_v$ and $v_v$ are the enthalpy and specific volume of vapour, $h_l$ and $v_l$ are the enthalpy and specific volume of liquid. In the incremental form, one gets :

$$dv = (v_v - v_l)d\chi + \left[\chi\left(\frac{\partial v_v}{\partial P}\right)_{Sat(T,P)=0} + (1-\chi)\left(\frac{\partial v_l}{\partial P}\right)_{Sat(T,P)=0}\right]dP \tag{10}$$

$$dh = (h_v - h_l)d\chi + \left[\chi\left(\frac{\partial h_v}{\partial P}\right)_{Sat(T,P)=0} + (1-\chi)\left(\frac{\partial h_l}{\partial P}\right)_{Sat(T,P)=0}\right]dP \tag{11}$$

where the partial derivatives are computed such that the pressure and temperature are such that increments are constrained to lie on the saturation curve:

$$dT = \left(\frac{\partial T}{\partial P}\right)_{Sat(T,P)=0} . dP \tag{12}$$

Relations (10,11) may be inversed such that $d\chi$ and $dP$ are expressed according to $dv$ and $dh$. Note that the partial derivatives in Eqs. (10,11) are computed from the tabulated values in Eqs. (5). For a given couple of volume and enthalpy at time $t^{(n)}$ denoted as $(v^n, h^n)$ and standing on the saturation curve, at the state n+1, the increments of volume and enthalpy are $\Delta h$ and $\Delta v$. The incremental enthalpy corresponds to the energy injected and the incremental specific volume is the solution of the continuity equation at state n. These increments are subdivided into $j$ sub-increments whose size corresponds to the increments in the tabulation of the thermodynamic variables. This is necessary for a proper evaluation of the various partial derivatives in Eqs. (10,11). Within each sub-increment, say sub-increment $k$, defined by $\Delta v^k$ $\Delta h^k$, the corresponding increments of pressure $\Delta P^k$ and the mass fraction of vapour $\Delta\chi^k$ are computed. It defines a new intermediate thermodynamic state corresponding to a pressure, specific volume, enthalpy, and mass fraction of vapour from which the next increment $\Delta v^{k+1}$ $\Delta h^{k+1}$ may start. Once all the sub-increments have been computed, and upon the condition that $\chi$ is comprised between 0 and 1, residual increments are calculated as:

$$\Delta v_r = v^n + \Delta v - \sum_{k=1,j} \Delta v^k \quad \text{and} \quad \Delta h_r = h^n + \Delta h - \sum_{k=1,j} \Delta h^k \tag{13}$$

When the residual increments exceed the accuracy criterion (**donner le nombre!!!)**, the calculation restarts with shorter time steps which corresponds to smaller increments. Note that is the mass fraction of vapour is reaches either 0 or 1, the same type of calculation is performed, with vapour or liquid water alone respectively.

*3.2 Energy injection*

The electrical discharge is described by the injection of energy into a mass of water between the electrodes. The zone where the energy is injected is a sphere with diameter equal to the inter-electrode distance.

The total injected energy is defined by:

$$W = \int_0^\infty \sum_{e=!}^{N} m_e^c \dot{q}_e(t)\, dt \tag{14}$$

$m_e^c$ is the mass of water in the element *e* where the energy is injected, it is assumed to be constant during the application of energy. *N* is the number of finite elements in the area where the energy is injected.

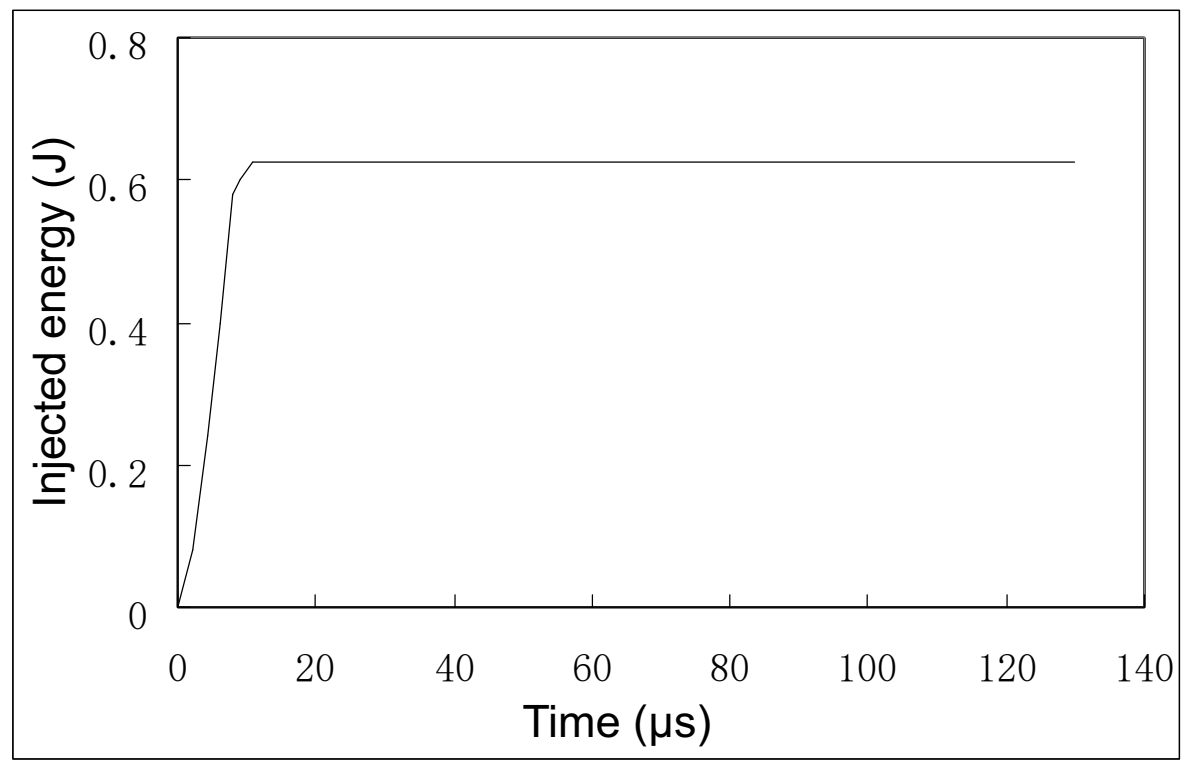

**Figure 3.** *Example of evolution of injected energy with time*

$\dot{q}_e(t)$ is the specific electrical power injected. It is computed according to the following formula [19]:

$$\dot{q}_e = \alpha_m \times C \times f(t) \tag{15}$$

where $f(t)$ is a function provided by the user which gives the variation in time of the power (see an example in Fig. 3), $C$ is a constant which may be changed from one element to another in order to account for a non homogeneous distribution of power in water (it is set constant in the present computations ), and $\alpha_m = 1$, which means that the injected energy is independent of the mass and the nature of the fluid.

*3.3 Finite element description*

During the underwater electrical explosion, dynamic and instability bubbles are formed [20]. These high pressure bubbles expand very quickly and move the boundary between water and the product gas at a high velocity. In the calculations, this phenomenon induces severe distortion and deformation of the mesh. The electrical discharge is difficult to model with a Lagrangian description. It is very difficult to obtain a steady state detonation owing to the motion of the grid at the wave velocity [21]. Hence, a perfect Euler grid system has been used in the numerical simulations. In the foregoing calculations, an axisymmetric model has been used, taking advantage of the specific configuration of the experimental set-up. The time integration scheme is explicit, with a damping that can be tuned in order to attenuate spurious oscillations.

**4. Results and comparisons with the experiments**Our objective is to achieve a calibration of the damping coefficient and of the injected power function in order to describe the profile and propagation of the shock wave induced by the electrical discharge. These parameters were chosen in order to obtain the numerical pressure v.s. time curves as close as possible to experimental results. In further calculations, the pressure v.s. time curves in different locations have been computed and also compared with experimental data. Two kinds of calculations have been carried out, with large and small electrodes respectively. In the first one, a great quantity of energy (3.3kJ) was injected with the help of large steel electrodes of diameter equal to 3cm. The steel electrodes were described as rigid bodies in order to observe the influence of the reflection of the pressure waves on the electrodes. The numerical parameters in the time stepping scheme have been fitted on this configuration. Two methods to inject energy are discussed. In the second kind of calculations, a low level of energy has been injected with small the steel electrodes of diameter equal to 0.5 cm. These little electrodes barely influence the propagation of pressure waves. The numerical pressure-time curves were also compared to experimental results.

### *4.1 Simulation with large electrodes*

The high quantity of energy (3.3kJ) is injected with large electrodes, assumed to be perfectly rigid. The diameter of electrode is 3cm and the inter-electrode distance is 1cm. The boundary conditions are described in Fig. 4. The characteristics used for the numerical model are: Water $\rho = 998.3 kg/m^3$, initial pressure $P^{(0)} = 10^5 Pa$ (atmospheric pressure) and initial temperature $T^{(0)} = 25^O C$.

*Figure 4. Boundary conditions of model with large electrodes*

Two functions have been used for the injection of electrical energy. In the first one (fig 5.a), the energy is injected into a 10mm diameter sphere (shown in yellow) between the two electrodes. We assume that when the injection begins, little bubbles were generated around the electrodes. If the injection time is long enough, these bubbles coalesce and form a large bubble which fills the entire inter-electrode space. Then, electrical breakdown occurs and the shock wave is generated. In the second model (fig 5. b), there are two steps in the injection process. In the first step, the energy is injected uniformly into a 10mm diameter sphere (shown in yellow) between the two electrodes until the temperature reaches 95C° in this zone. This step simulates the heating stage, prior to bubble formation. Then, the rest of the energy is injected into a small strip (shown in red) which is at the centre of the sphere and parallel to the axis. Here, we assume that the spherical zone is heated but the bubbles form only into a small part of the inter-electrode space where the electrical discharge occurs.

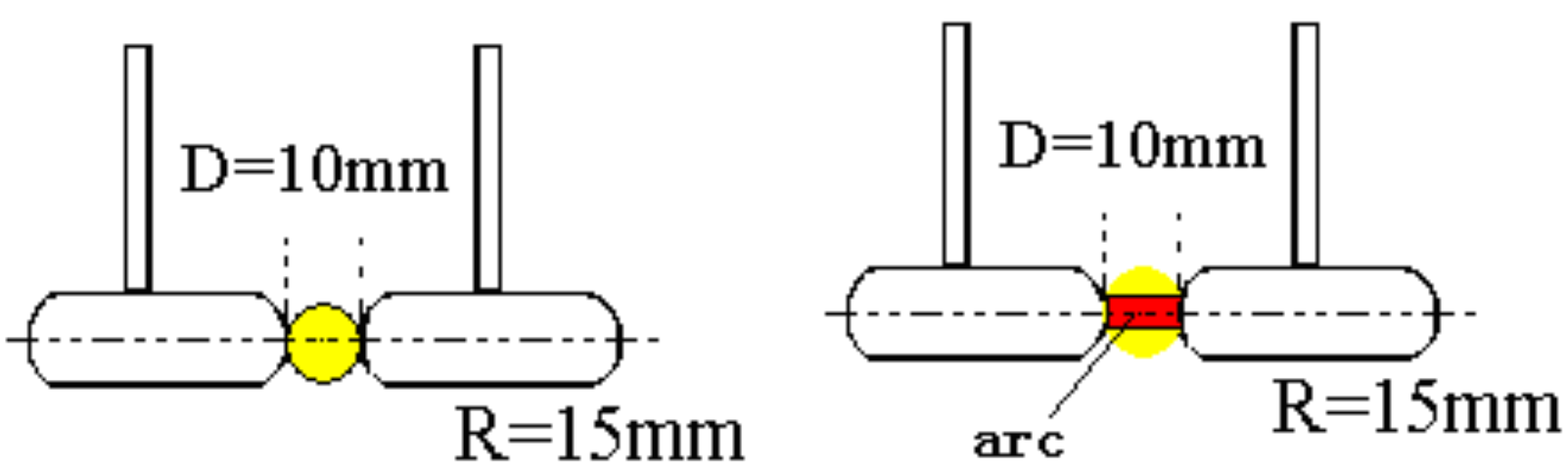

*Figure 5. First (a) and the second (b) energy injection modes*

The time step in the explicit time integration is provided by the classical CFL condition. Hence, the mesh should be as uniform as possible (Fig. 6). This mesh is as fine as possible in order to capture the pressure wave as accurately as possible.

*Figure 6. Mesh of the first model in the simulation with electrodes*

Figure 7 presents the mesh which has been used in the second mode. This mode includes two steps in the energy injected process. In the first step, energy was injected uniformly into a sphere zone between the two electrodes until the maximum temperature in this zone reached 95°C. This heating process consumed 328 Joules, which was about 10% of the total energy involved in the process. Then, the rest (90% of the energy) was injected into the little strip (shown in red in figure 7) in order to simulate the electrical arc.

**Il faut donner ici les profils d'énergie injectée dans les deux modèles!!!!!!!!!!!! Avec des commentaires sur la façon dont ils ont été obtenus…..**

*Figure 7. Mesh of the second mode in the simulations with the electrodes*

Damping is very important in such simulations. It has a large influence on the profile of the pressure wave. In this explicit calculations, the damping coefficient is introduced in order to attenuate high frequency oscillations generated by the discretization. On the other side, damping should not attenuate the pressure waves, and therefore should be as small as possible. **Ici il faut faire un rappel des principales equations pour montrer ou le damping intervient. Montrer l'effet de la variation du damping et que le premier mode est meilleur??????? Donner enfin les valeurs du damping utilisées.** The comparison between the experimental and numerical pressures at a point located at a horizontal distance of 17.5cm from the axisymmetry axis, on the horizontal axis of symmetry is shown in Fig. 9. The peak pressure of the second mode is lower than the one of the first mode, which is closer to the experimental result.

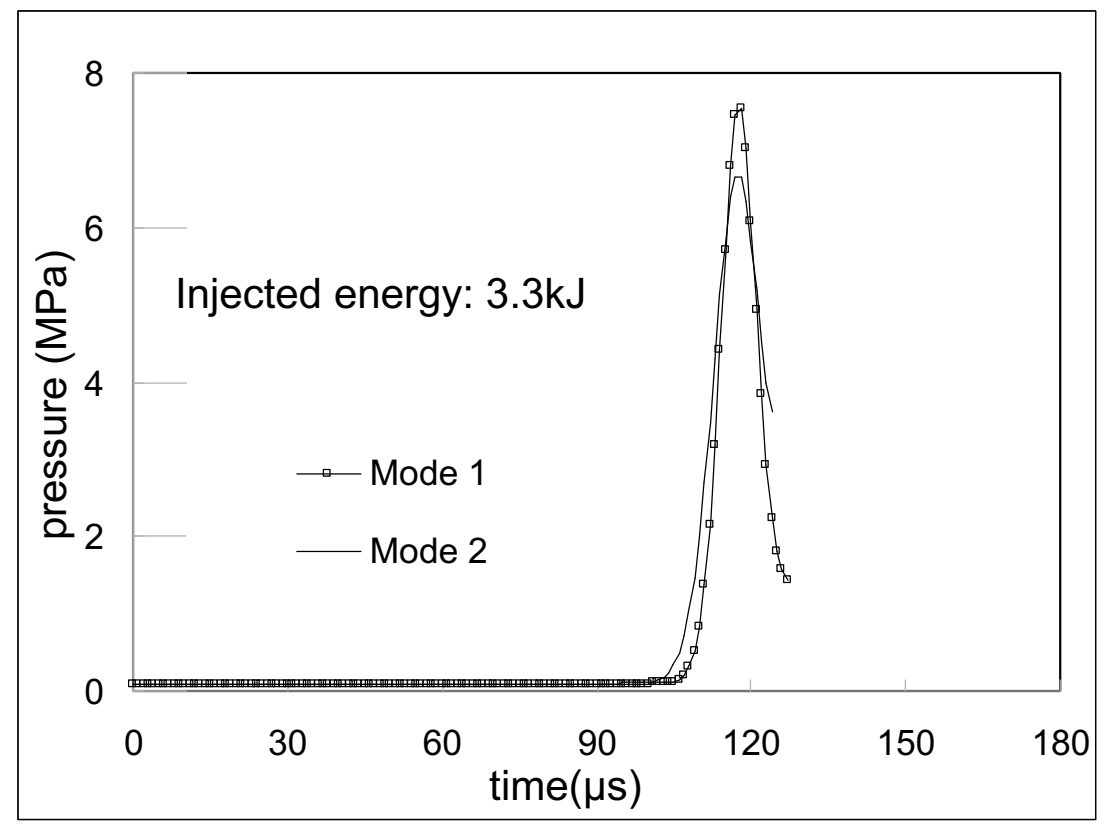

*Figure 8. Comparison of the numerical pressures*

In order to examine the capabilities of the numerical model, we have compared the computed peak pressures at different horizontal distances from zone where the electrical discharge occurs (see Figs 6,7) with experimental data. The experimental data have been used to fit Eq. (3) The empirical formula holds for distances to the arc larger than 5 cm approximately as measurements could not be performed closer. Fig. 9 shows that there is a good agreement between experimental data and the calculations. , There is no distinct difference between the two numerical models considered. Since the second one exhibits larger oscillations than the first one which is better as seen in Fig. 8, we are going to consider only this model in further calculations.

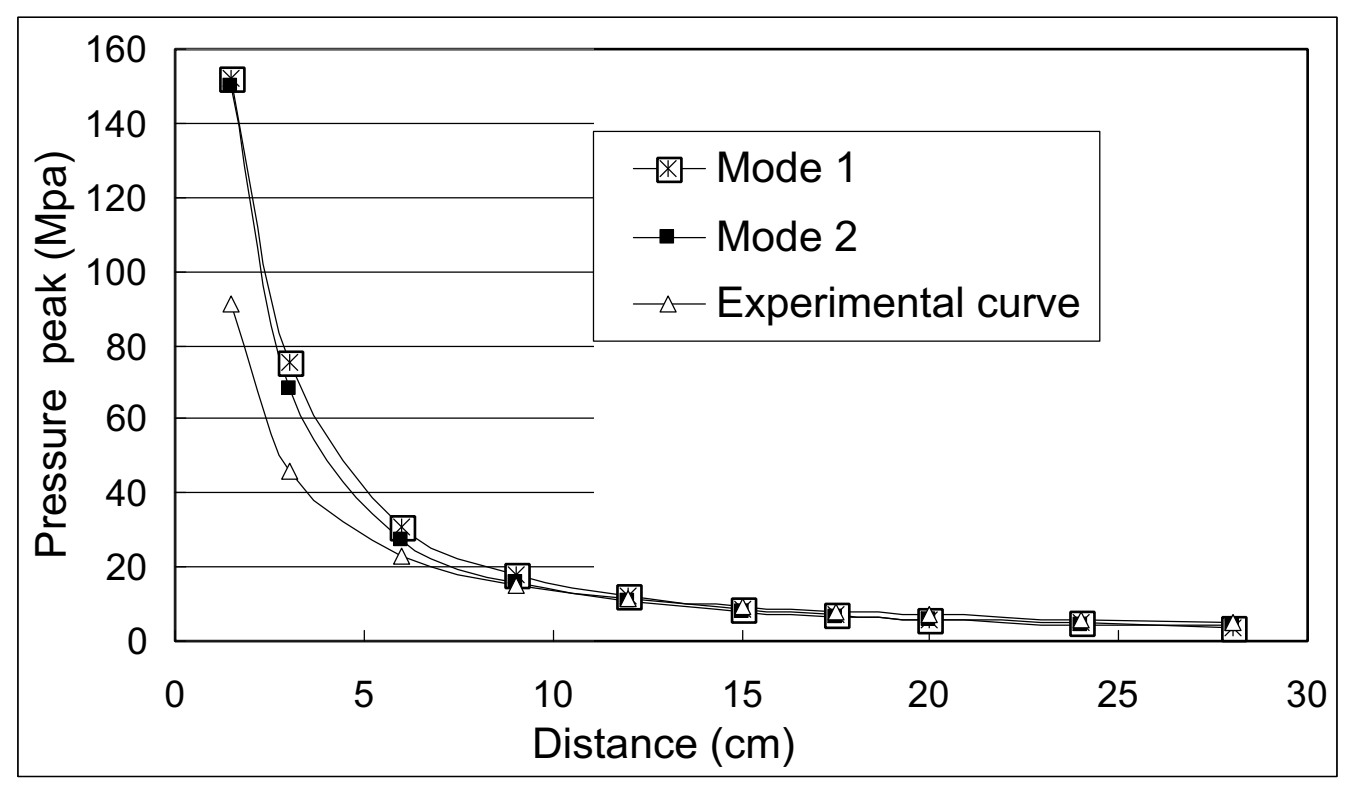

*Figure 9. Comparison of the numerical peak pressures relative to the distance between the measured point and the charge with the empiric formulation*

*4.2 Simulations with small electrodes*

In this series of simulation, a uniform mesh was used and the electrodes were ignored due to their small size. Since the injected energies ire very small, the regions where energies were injected should be rery small too. Hence, the mesh must be very fine to match the numerical accuracy. The finite element mesh is shown in Fig.10. T has been used in the follo

Three examples are treated corresponding to three experiments respectively. In the first one, the injected energy is 600.625J and the pressure is measured at a point situated at 17.5cm from the electrodes. In the second and third examples, the pressure is measured at a distance of 9cm from the electrodes; the injected energies are 31.25J and 20 J respectively.

Electrical energy was injected into a zone (shown in red) the volume of which was calculated from the inter-electrode spacing. As is mentioned in foregoing section, about 10% of total energy was used to heat water. So, the thermal energy in the formulation was supposed to be 10% of total injected energy. In the three examples, the volumes of the injected zone are 205.2mm$^3$, 10.6mm$^3$ and 6.8mm$^3$ respectively. The simulations used the same boundary condition and characteristics as in the previous series of computations.

*Figure 10. The mesh in the second series of simulations*

The three pressures histories are compared with the experimental results respectively in Fig 11. In the Fig 11 a), the experimental and numerical curves present pressure peaks of 44.076MPa and 42.448MPa respectively, the relative error equals 3.6%. In Fig 11 b), the pressure peak of the experimental and numerical curves are 28.694MPa and 26.429MPa, the relative error equals 7.8%. Fig 11 c) shows the experimental and numerical curves with the pressure peak of 20.2MPa and 21.8MPa respectively, the relative error is 7.3%. The figure 12 illustrates the variation of numerical pressure and experimental results with injected energy. The experimental data have been complemented with those from Touya et al [4]. A consistent agreement between the simulation and experimental data is observed.

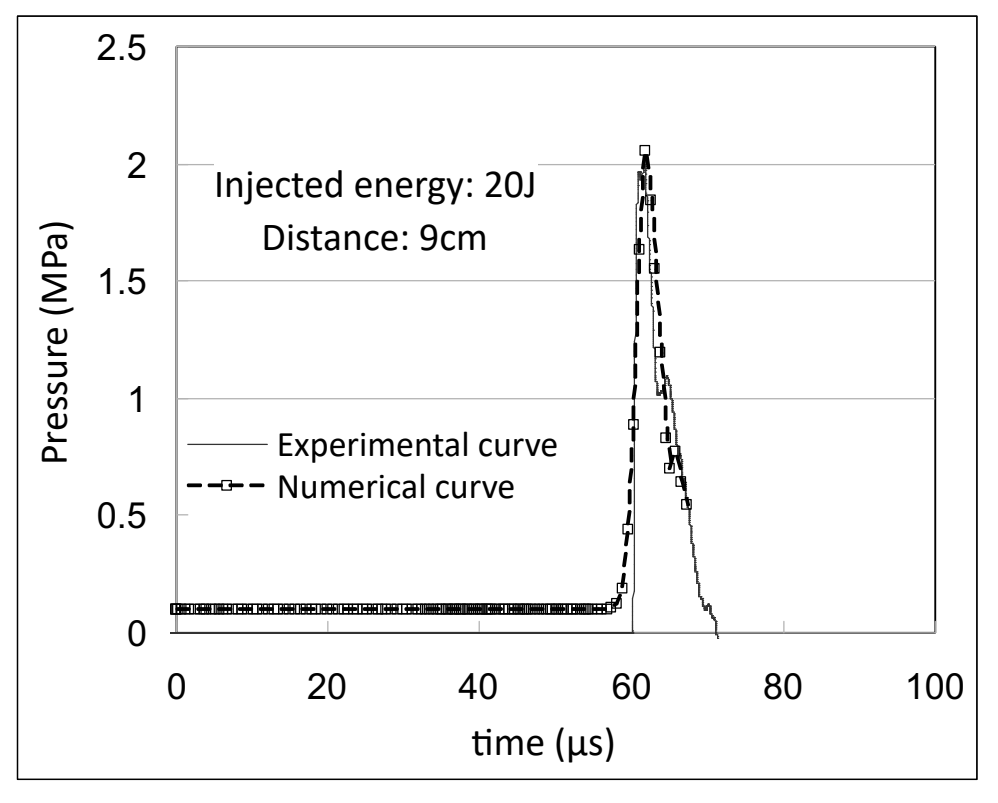

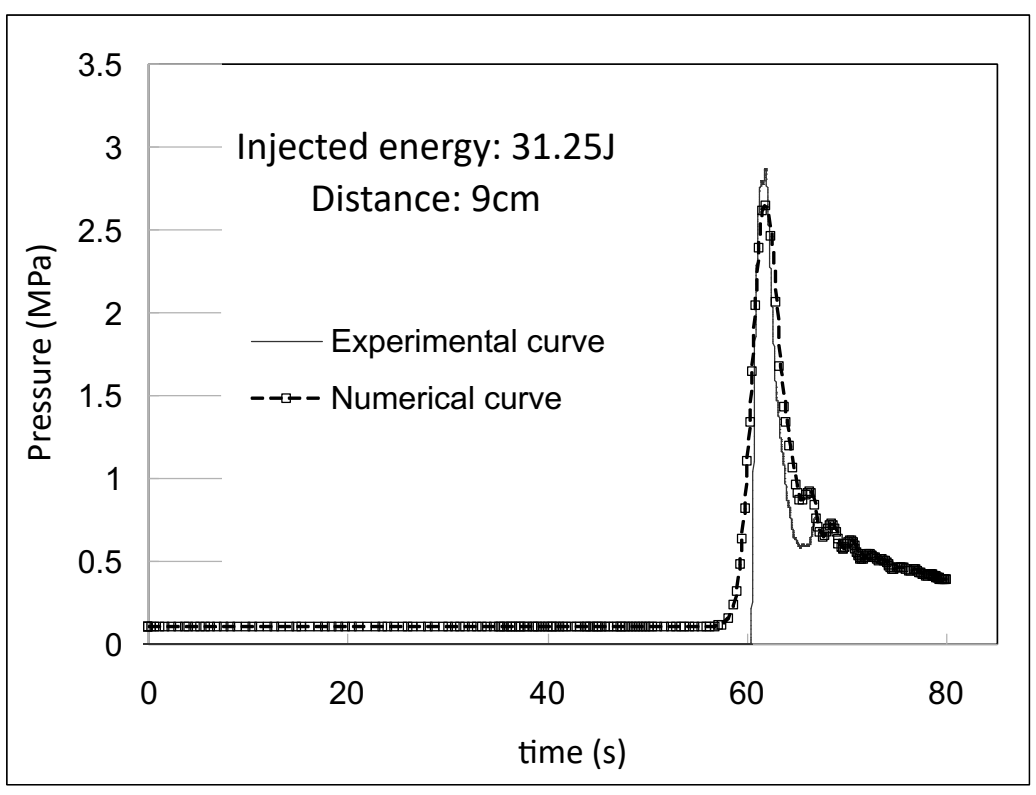

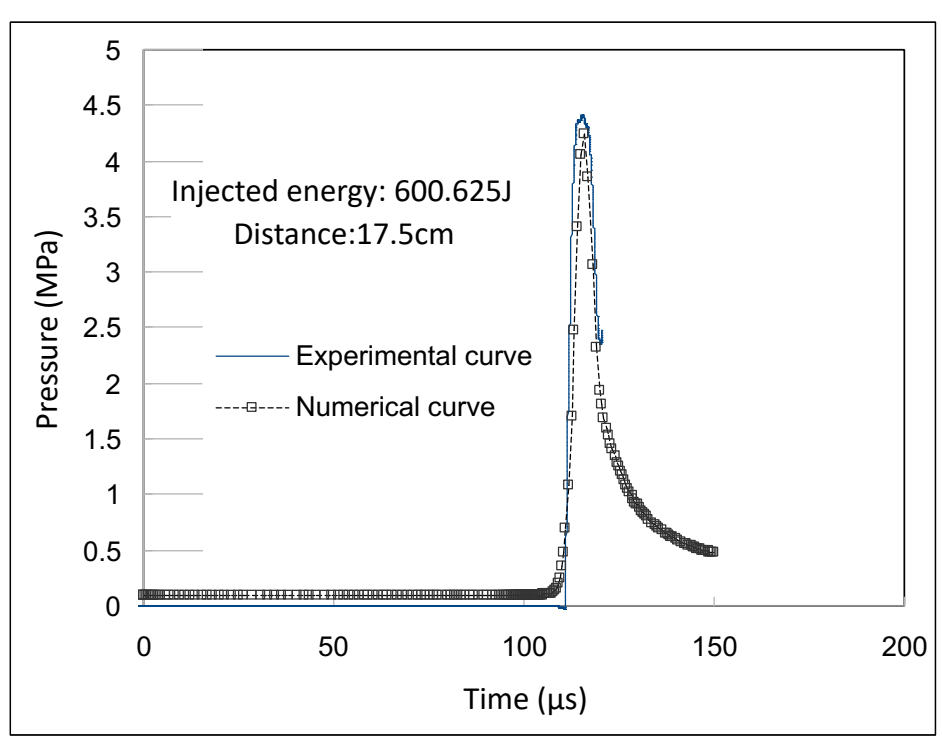

*Figure 11. Comparison of the experimental and the numerical pressures*

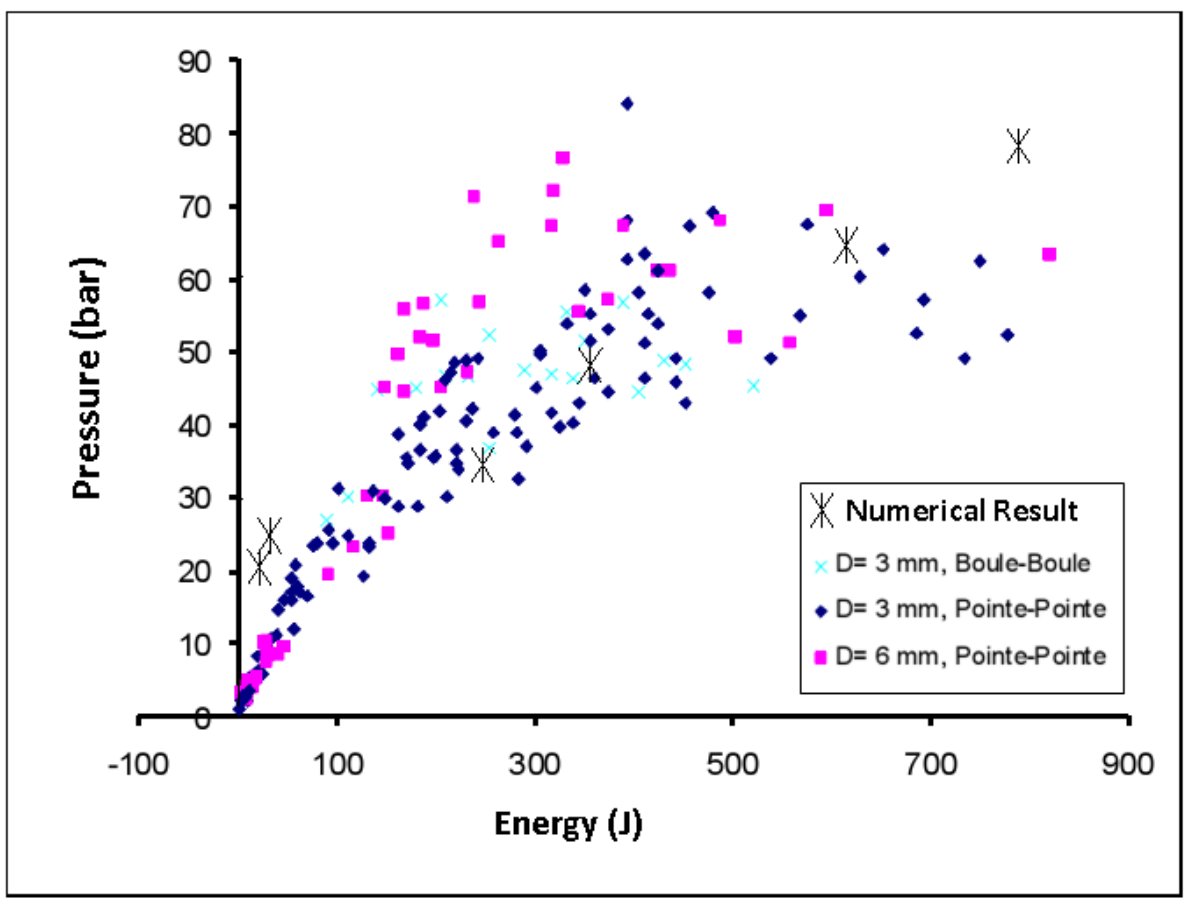

*Figure 12. Comparison of the experimental [4] and the numerical pressures at 9cm*

## 5. Conclusions

This paper deals with the development of a simplified method to simulate the propagation of the shock wave in water due to an explosion. The blast is a shock wave generated in water by Pulsed Arc Electrohydraulic Discharges (PAED). Since few experimental data were available in the scientific literature, an experimental program has been carried out. Four injected energy have been used. Two different sensors have been used to register the pressure waves at two positions respectively.

The generation of shock wave is simulated by a simplified method without taking into account of the physical parameter distributions in the plasma channel. The electrical discharge is modeled as procedure of energy injection. The injected electrical energy is simulated as augmentation of enthalpy in local water zone. Numerical simulation of experiments was carried out using EUROPLEXUS for rapid dynamic response of fluid to compare with the experiments' results. An Eulerian formulation was used in this simulation. Two different kinds of calculation have been carried out with big and small electrodes respectively. In the first calculation, the big electrodes were not simulated but represented by a fixed curve. In the second calculation, the small electrodes were not taken account. Good agreements for the peak pressure, the duration of pressure pulse and the reach time with experimental data are obtained. The simulation of the propagation of pressure waves in water and structure is was presented in this paper as well. The damage due to the pressure waves is obtained by a damage concrete model. The anisotropic permeability has been calculated by the principal damage. The good correlation between the experimental and numerical results represents a good approximation for this problem.